\documentclass[prx,superscriptaddress,amsmath,amssymb,twocolumn,longbibliography]{revtex4-1}
\usepackage{graphicx}
\usepackage{subfigure}
\usepackage{adjustbox}
\usepackage{bm}
\usepackage{color}
\usepackage{braket}
\usepackage{standalone}
\usepackage{multirow}
\usepackage{tikz}
\usepackage{enumitem}
\usepackage{mathrsfs}
\usepackage{dsfont}
\usepackage{comment}
\usepackage{amsmath}
\usepackage{lipsum}
\usepackage[colorlinks,bookmarks=true,citecolor=blue,linkcolor=red,urlcolor=blue]{hyperref}
\usepackage[capitalize]{cleveref}
\usepackage{soul}

\usepackage{tikz}
\usetikzlibrary{math}

\newcommand{\be}{\begin{equation}}
\newcommand{\ee}{\end{equation}}

\newcommand{\ba}{\begin{eqnarray}}
\newcommand{\ea}{\end{eqnarray}}

\newcommand*{\id}{{\normalfont\hbox{1\kern-0.15em \vrule width .8pt depth-.5pt}}}

\newcommand{\commentblock}[1]{}

\graphicspath{{./figures/}}

\begin{document}

\title{Variational Quantum Simulations of a Two-Dimensional Frustrated Transverse-Field Ising Model on a Trapped-Ion Quantum Computer}
\author{Ammar Kirmani}
\email{akirmani@lanl.gov}
\affiliation{Theoretical Division, Los Alamos National Laboratory, Los Alamos, New Mexico 87545, USA}

\author{Elijah Pelofske}
\email{epelofske@lanl.gov}
\affiliation{Information Systems \& Modeling,
Los Alamos National Laboratory, Los Alamos, New Mexico 87545, USA}

\author{Andreas B\"{a}rtschi}
\email{baertschi@lanl.gov}
\affiliation{Computer, Computational and Statistical Sciences Division,
Los Alamos National Laboratory, Los Alamos, New Mexico 87545, USA}

\author{Stephan Eidenbenz}
\email{eidenben@lanl.gov}
\affiliation{Computer, Computational and Statistical Sciences Division,
Los Alamos National Laboratory, Los Alamos, New Mexico 87545, USA}

\author{Jian-Xin Zhu}
\email{jxzhu@lanl.gov}
\affiliation{Theoretical Division, Los Alamos National Laboratory, Los Alamos, New Mexico 87545, USA}

\date{\today}

\begin{abstract}
Quantum computers are an ideal platform to study the ground state properties of strongly correlated systems due to the limitation of classical computing techniques particularly for systems exhibiting quantum phase transitions. While the error rates of Noisy Intermediate-Scale Quantum (NISQ) computers are still high, simulating strongly correlated systems on such devices and extracting information of possible phases  may be within reach. The frustrated transverse-field Ising model (TFIM) is such a system with multiple ordered magnetic phases. In this study, we simulate a two-dimensional frustrated TFIM with next-nearest-neighbor spin-exchange interactions at zero temperature. The competition between the nearest-neighbor ferromagnetic and next-nearest-neighbor antiferromagnetic coupling gives rise to frustration in the system. Moreover, the presence of quantum fluctuations makes the ground-state phase profile even richer. We use the Variational Quantum Eigensolver (VQE) to compute the phases on a square lattice with periodic boundary conditions for a system of $16$ sites (qubits). The trained VQE circuits are compared to exact diagonalization, allowing us to extract error measures of VQE. We focus on the ground-state phase transitions of this model, where VQE succeeds in finding the dominant magnetic phases. The optimized VQE circuits are then executed on the Quantinuum H1-1 trapped-ion quantum computer without using any error mitigation techniques. Our experiments show near perfect recovery of the magnetic phases of the frustrated model through ground-state energy, the energy derivative, and the spin correlation functions. Thus, we show  that the trapped-ion quantum processor is able to achieve reliable simulations of a strongly correlated system within the limitations of the VQE approach.
\end{abstract}

\maketitle

%%%%%%%%%%%%%%%%%%%%%%%%%%%%%%%%%%%%%%%%%%%%%%%%%%%%%%%%%%%%%%%%%%%%%%%%%%%%%%%%%%%%%%%%
\section{Introduction}
\label{section:introduction}
%%%%%%%%%%%%%%%%%%%%%%%%%%%%%%%%%%%%%%%%%%%%%%%%%%%%%%%%%%%%%%%%%%%%%%%%%%%%%%%%%%%%%%%%
Simulating the properties of strongly-interacting quantum systems is of considerable interest in physics and chemistry for systematically understanding exotic physical phenomena. An important limitation, however, is that there do not exist efficient classical algorithms that can simulate such systems in particular for large system sizes at zero temperature, where quantum effects are most dominant~\cite{PhysRevE.98.033309, iskakov2018exact, lauchli2010numerical}. Some classical-based methods can simulate a two-dimensional (2D) quantum system at zero-temperature by solving a 2+1-D classical system for finite $T$, thus adding further complexity to the problem~\cite{2D3D}. Quantum computing is a potentially viable alternative route to compute the properties of interacting quantum-mechanical systems and in particular, magnetic quantum systems~\cite{King_2021, Zhang_2012, Struck_2011, King_2021_scaling, prrbenedikt}. The current era of Noisy Intermediate Scale Quantum (NISQ) devices presents several challenges e.g., the presence of noise in state preparation, readout errors, and crosstalk etc. Earlier studies have shown that quantum simulations of correlated systems can be improved through the use of error mitigation methods~\cite{prrbenedikt, ibmeagle}.

The problems naturally suited for quantum computers include ground state preparation and out-of-equilibrium time-dependent phenomena of strongly correlated quantum systems. For the ground state preparation, many quantum algorithms have been proposed e.g., adiabatic state preparation, variational quantum algorithms (VQAs)~\cite{natureCerezo}, quantum phase estimation (QPE)~\cite{kang2022optimizedquantumphaseestimation, QPEIEE} and imaginary time-evolution methods~\cite{quantumimag}. VQAs, which are hybrid quantum-classical algorithms have emerged as the most promising algorithmic methods to leverage NISQ devices by keeping the noise levels low. VQAs execute Parameterized Quantum Circuits (PQCs) on quantum computers while outer-loop parameter optimization is performed on a classical computer. This approach keeps the quantum circuit depth shallow and is therefore well-suited for near-term NISQ devices~\cite{natureCerezo}. The Variational Quantum Eigensolver (VQE)~\cite{Peruzzo_2014, McClean_2016, Tilly_2022} is a type of VQA, which aims to find the ground state of a given Hamiltonian by optimizing an upper bound on the ground state energy ~\cite{VQEreview}. This method has been used to simulate properties of many different types of quantum systems~\cite{Kandala_2017, PhysRevA.104.L020402, PhysRevB.106.214429, PRXQuantum.2.020310}.

Several previous studies have examined the use of NISQ-friendly~\cite{Preskill_2018} quantum algorithms in order to simulate the physical properties of spin systems~\cite{10313877, PhysRevResearch.5.043217, PhysRevA.102.012415, PhysRevB.102.075104, PhysRevA.104.L020402, PhysRevB.106.214429}, in particular at or near phase transitions of frustrated magnetic systems~\cite{prrbenedikt, pexe2024usingfeedbackbasedquantumalgorithm, rattighieri2025acceleratingfeedbackbasedquantumalgorithms, guo2023performancevqephasetransition, Lukas_Bosse_2024, Zhang2023simulatinggauge}. The transverse-field Ising model (TFIM) is an example of such a system. At $T=0$ K, this model is exactly solvable in one dimension (1D)~\cite{TFIExact} and serves as benchmark for the performance of near-term quantum devices. Quantum simulation of this system at finite temperature shows for the first time the evidence of phase transitions in 1D quantum system~\cite{Monroe1d} when long-range interactions are considered. The dynamical properties of TFIM were addressed by proposing a variational based quantum circuit~\cite{gover2025fullyoptimisedvariationalsimulation}. Quantum simulations of Heisenberg spin chain in 1D has also been performed using Trotterization-based methods~\cite{chowdhury2024}. A hybrid framework combining variational circuits with classical machine learning models for detecting phase transitions in TFIM has been proposed recently~\cite{Cao_2025}. Large-scale simulations of a type of hardware-compatible TFIMs were successfully performed on IBM superconducting-qubit NISQ computer~\cite{ibmeagle} with the extensive use of error mitigation techniques. Although this simulation could not be classically replicated using exact state vector simulations, it was classically simulable using a variety of techniques~\cite{kechedzhi2023effective, begušić2023fast, tindall2023efficient, begušić2023fast_2, liao2023simulation, rudolph2023classical, patra2023efficient, anand2023classical, 2308.01339} in large part due to the short depth needed for the simulations and the sparse interaction graph of the Hamiltonian. 
The ground state phase diagram of a 1D frustrated TFIM has recently been studied~\cite{prrbenedikt} by using a short depth VQE hardware-efficient-ansatz~\cite{Leone2024practicalusefulness} (compatible with the hardware's heavy-hex graph) on superconducting-qubit IBM quantum processors. Due to prohibitively high error rates, many simple quantities like ground state energy did not reproduce qualitative agreement with exact results. Error mitigation techniques did not improve the situation to a large extent. Therefore, only after identifying noise-robust quantities associated with the magnetic phases of the model, the qualitative phase diagram was obtained ~\cite{prrbenedikt}. 

In this work, we consider a 2D TFIM on a square lattice with both nearest- and next-nearest-neighbor competing exchange interactions. To simulate the ground state properties of the model, we use the approach of Variational Quantum Eigensolver (VQE)~\cite{Peruzzo_2014, McClean_2016, Tilly_2022}. This is done by splitting the task of solving an eigenvalue problem into two parts: one part is a VQE parameter optimization task which is performed on a classical computer, and the second part is evaluating an observable of a quantum state, usually the energy of a Hamiltonian using a (noisy) quantum computer. Given a  relatively short-depth parameterized quantum circuit, the goal of VQE is to train the parameterized circuit by using a classical optimization algorithm~\cite{Peruzzo_2014, McClean_2016, Tilly_2022} to approximate a quantum state of interest (typically ground state(s)). Strikingly, even without invoking quantum error mitigation on the trapped-ion quantum device, we are able to identify different magnetic phases in a parameter space spanned by the magnetic field and competing magnetic exchange interactions. The VQE results are well benchmarked by exact-diagonalization results.

The remainder of the paper is organized as follows: In Sec.~\ref{section:model}, we define our model with possible ordered phases. Sec.~\ref{section:methods} gives the details of methods used in this work. Results of quantum simulations carried out on a Quantinuum trapped-ion quantum computer are given in Sec.~\ref{section:results}, followed by a discussion in Sec.~\ref{section:conclusion}.

\section{Model}
\label{section:model}
The transverse field Ising model on a 2D square lattice with additional frustrating next-nearest-neighbor interactions, as shown in Fig.~\ref{fig:VQE_circuit_ansatz}(a), is given by. 
\begin{align}
\label{eq:Hamiltonian}
{\mathcal H} = -J_1 \sum_{\langle i,j\rangle} Z_i Z_j + J_2 \sum_{\langle\langle i,j \rangle\rangle} Z_i Z_j + B_x \sum_i X_i\;, 
\end{align}
where $X_i$, $Z_i$ are on-site (qubit) Pauli operators $i\in [0,N-1]$ and $N$ is the system size. $\langle ... \rangle$ ($\langle\langle ... \rangle\rangle$) denotes that the sum is restricted to nearest neighbor (next-nearest neighbor) pairs of the spin exchange interaction on the square lattice. In our notation, positive $J_1$ ($>0$) enforces ferromagnetic coupling between the spins while positive $J_2$ ($>0$) denotes antiferromagnetic coupling. Hereafter we set $J_1=1$ and vary both next-nearest-neighbor interaction $J_2$ and the transverse field strength $B_x$ as control parameters. Due to the competition between nearest-neighbor and next-nearest-neighbor interactions and the presence of the quantum fluctuations induced by the transverse field $B_x$, the system exhibits a rich ground-state profile. This model has been previously studied, such as in Refs.~\onlinecite{PhysRevE.97.022124, Oitmaa_2020_TFANNNI}. In Ref.~\cite{Oitmaa_2020_TFANNNI}, using series expansion, the ground-state phase profile is obtained as frustration parameter ($J_2/J_1$) and transverse field is changed. Similarly, a frustrated model for the antiferromagnetic nearest-neighbor case has also been considered~\cite{pre_anti2D}. In the absence of transverse field and finite temperature, this model can be studied by Monte Carlo based methods~\cite{pre_TANNNI_classical}. 

The ground-state phase diagram of this Ising model (even in 1D) is still a subject of debate~\cite{prrbenedikt,phase1d}, where there appears to be no consensus on the number of phases or the exact location of phase transitions. Although, one needs the system in the thermodynamical limit to characterize the ground-state phase profile, methods like cluster mean-field~\cite{pre_anti2D}, density-matrix renormalization group~\cite{PrrDMRG}, quantum Monte-Carlo methods~\cite{MisawaMonte} and perturbative based methods~\cite{Oitmaa_2020_TFANNNI} point to the possibility of certain phases. For $J_2=B_x\to0$, the system prefers ferromagnetic ordering while for $J_2/J_1\approx1$ and $B_x\to 0$, the system has a stripe-ordered ground state, where spins are anti-parallel in two consecutive columns (rows) of the lattice as shown in Fig.~\ref{fig:VQE_circuit_ansatz}(b). For large values of $B_x$, the system exhibits a field-induced spin-polarized state.

%%%%%%%%%%%%%%%%%%%%%%%%%%%%%%%%%%%%%%%%%%%%%%%%%%%%%%%%%%%%%%%%%%%%%%%%%%%%%%%%%%%%%%%%
\section{Methods}
\label{section:methods}
%%%%%%%%%%%%%%%%%%%%%%%%%%%%%%%%%%%%%%%%%%%%%%%%%%%%%%%%%%%%%%%%%%%%%%%%%%%%%%%%%%%%%%%%
We apply the Variational Quantum Eigensolver algorithm to perform quantum simulations of the 2D frustrated TFIM given in Eq.~\eqref{eq:Hamiltonian} with periodic-boundary conditions (see Appendix~\ref{section:appendix_periodic_boundary_condition}). Different quantum-computing architectures offer different qubit connectivity ranging from superconducting-qubit quantum computers usually offering limited connectivity to trapped-ion computers offering all-to-all connectivity. Limited connectivity in the case of superconducting-qubits can be overcome by adding more entangling gates between non-connected qubits which ultimately degrades the performance of the system ~\cite{MonroeComparison}. Moreover, the presence of periodic-boundary conditions further complicates the implementation of quantum algorithms on limited-connectivity architectures. Thus, our 2D system is especially well suited for the all-to-all connectivity of trapped-ion quantum computers. 

\begin{figure*}[th!]
    \centering
    \includegraphics[width=1.0\textwidth]{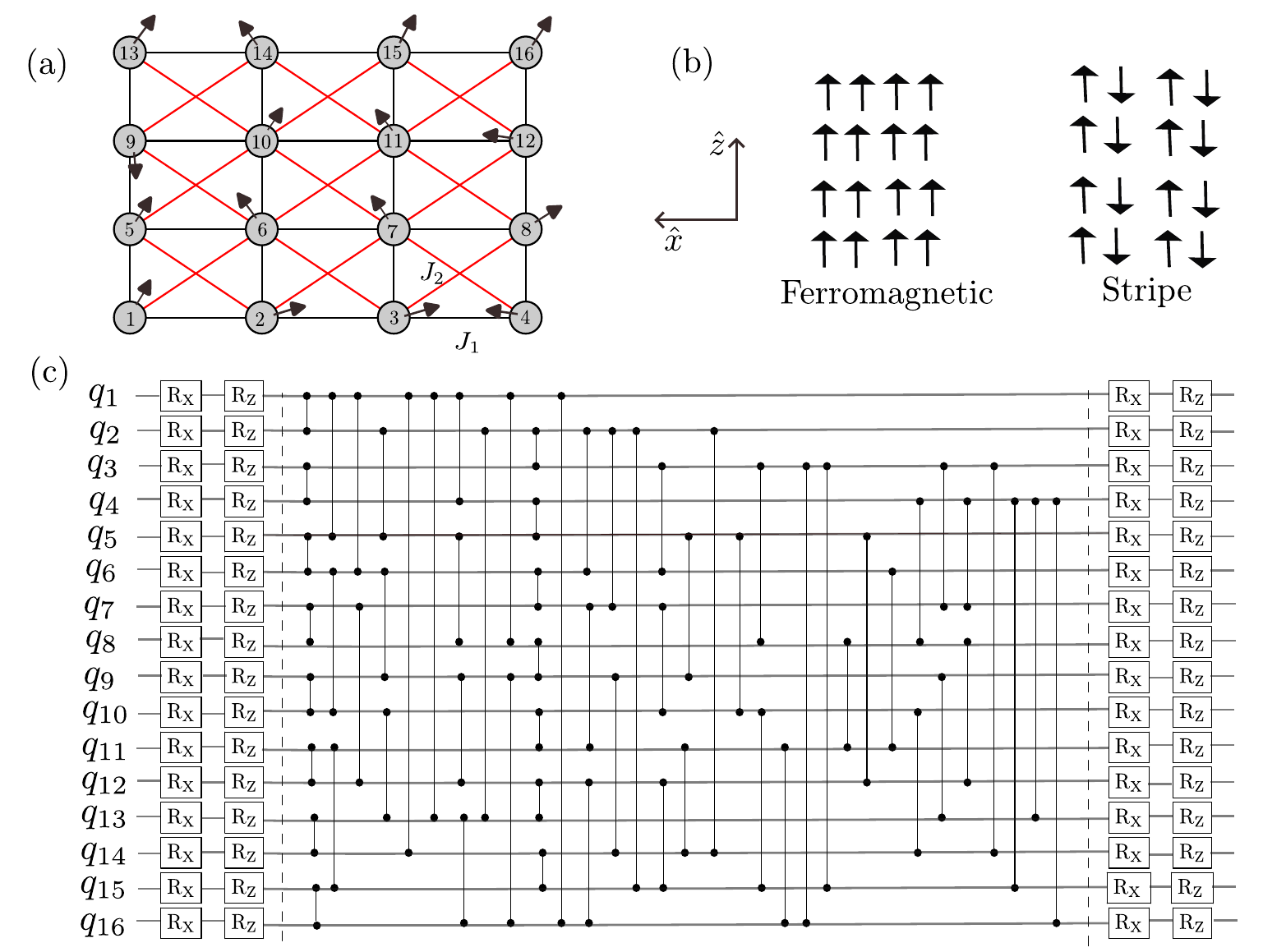}
    \caption{ (a) 2D transverse field Ising model with $J_1$ (nearest-neighbor) and $J_2$ (next-nearest-neighbor) interactions, shown by black and red bonds respectively. The transverse field is along $\hat{x}$ direction. The arrows represent spins that can take any direction in z-x plane. (b) Spin orientations of ferromagnetic and stripe-phase instability. (c) VQE ansatz for a $4\times4$ qubit system. Each qubit has 4 single-qubit (\texttt{Rx} and \texttt{Rz}) gates that are parameterized by 4 independent variational parameters with a total of $16\times 4$ parameters for the whole circuit. The entangling layer is composed of all CZ gates; one CZ gate for each spin-spin interaction term in Eq.~\eqref{eq:Hamiltonian}. Since the CZ gates commute with each other, their execution order in the entangling layer of (c) does not affect quantum state that is prepared. The state of all $16$ qubits is measured at the end of the circuit in the Z-basis (measurement operations are omitted to conserve space). }
    \label{fig:VQE_circuit_ansatz}
\end{figure*}

We first label the qubits according to our square-lattice model where each qubit corresponds to a particular lattice site (see Fig.~\ref{fig:VQE_circuit_ansatz}(a) and~\ref{fig:VQE_circuit_ansatz}(c)). The VQE circuit utilizes an all-to-all connectivity and is comprised of several layers of single-qubit and entangling gates. The first layer of the circuit has two single-qubit rotation gates on each qubit. The output of each of the rotation gate is controlled by a single continuous parameter. It is followed by an entangling layer of CZ gates corresponding to the two-site interactions among nearest and next-nearest-neighbors bonds contained in the Hamiltonian of Eq.~\eqref{eq:Hamiltonian}. The last layer contains another sequence of single-qubit rotation gates.
\onecolumngrid

\twocolumngrid
 Fig.~\ref{fig:VQE_circuit_ansatz}(c) gives the VQE circuit diagram that is used throughout this work for the $N=16$ site system. The order of the CZ gates does not alter the quantum state that is prepared by the ansatz because CZ gates commute with each other, in the case of the entangling layer in Fig.~\ref{fig:VQE_circuit_ansatz}(c), the order of the CZ gates is arbitrary. This VQE circuit can be effectively considered a type of hardware-efficient ansatz for a trapped-ion quantum computer because of the all-to-all connectivity between the qubits. The circuit effectively creates a unitary $\hat{U}(\boldsymbol{\theta})$ with continuous parameters $\boldsymbol{\theta}=(\theta_1,\theta_2,...,\theta_{4N})$. We then apply this unitary to obtain quantum state as $\ket{\Psi(\boldsymbol{\theta})}=\hat{U}(\boldsymbol{\theta})\ket{0}^{\otimes N}$. $\ket{\Psi(\boldsymbol{\theta})}$ is then optimized to give 
\begin{equation}\label{eq:costfuntion}
E(\boldsymbol{\theta})=\bra{\Psi(\boldsymbol{\theta})}{\mathcal H}\ket{\Psi(\boldsymbol{\theta})}\geq E_{GS} \;,
\end{equation}
where $E_{GS}$ is the ground state of the quantum Hamiltonian in Eq.~\eqref{eq:Hamiltonian}. We can map-out the ground-state phases for a given $B_x$ and frustration parameter $J_2/J_1$ once a reasonably accurate-upper bound $E(\boldsymbol{\theta})$ to $E_{GS}$ is found by classical optimization of the parameterized VQE circuit. The NISQ computer that we use is the Quantinuum H1-1 trapped-ion processor~\cite{Pino_2021}\footnote{The H1-1 quantum computer was used between February and April 2025.}. This quantum computer has a fully connected two-qubit gate connectivity, and has $20$ operational qubits which is sufficient for the $16$ qubit model we consider in this study. The VQE circuits are compiled using pytket and circuit optimization level of $2$ was chosen ~\cite{cowtan_et_al, sivarajah2020t} (which is the maximum, and the default). The hardware native (universal) gateset of the H1-1 trapped ion processor is $U_{1q}(\theta, \phi) = e^{\frac{-i\theta}{2}(\cos(\phi) X + \sin(\phi) Y)}$, $RZZ(\theta) = e^{-i \frac{\theta}{2} Z \otimes Z}$, $R_Z(\lambda) = e^{-i\frac{\lambda}{2}Z}$ and a general SU(4) entangling gate ($\theta$, $\lambda$ and $\phi$ denote the tunable parameters for the gates). Other gates can be expressed in the terms of the native gates such as $ZZ =RZZ(\pi/2)$, $R_X(\lambda)=e^{-i\frac{\lambda}{2}\hat{X}}=U_{1q}(\lambda,0)$ and $CZ=[I\otimes R_z(-\pi/4)]ZZ[I\otimes R_z(\pi/4)]$. 

In previous studies, local minima are avoided during the optimization by changing the VQE cost-function to the overlap of state generated by the variational circuit with classically pre-computed exact ground states~\cite{prrbenedikt}. To have a more comprehensive understanding of the VQE optimization loop, we compute the result of the parameterized VQE circuit using many classical blackbox learning iterations with random initialization of the parameters. We then select the VQE parameters yielding the lowest Hamiltonian expectation value found across all learning iterations to execute on the noisy quantum computer and quantify its output signal. The phases of the model are then obtained from this experimental signal measured on the quantum computer. The classical training of VQE serves two purposes. The first is that it reduces shot-count, and therefore total computation time, used by the quantum computer. The second is that it allows a clear ideal-case study to answer how well VQE can perform if trained using a classical-quantum feedback loop. 

To evaluate the expectation value of the Hamiltonian in Eq.~\eqref{eq:Hamiltonian}, we need to perform measurements in two different Pauli bases $Z$ and $X$. In particular, the $Z$-basis quantities such as the expectation values of nearest-neighbor bonds are extracted by measuring the system in the computational basis, which is the $Z$-basis. For the $X$-basis measurements (needed to evaluate the expectation of $B_x\sum_i X_i$), the qubits are rotated into the $X$-basis by applying a Hadamard gate to every qubit at the end of the VQE circuit, followed by a measurement of all qubits in the computational basis. Thus for each value of $B_x$ and $J_2$ considered in this study, we need to run the circuit (with the same learned VQE parameters) twice on the quantum computer (or simulator) in order to estimate the expectation value of the Hamiltonian. When executing VQE parameter training, a finite number of samples (shots) is used - specifically $10^4$ samples are computed for each Pauli-basis measurement of the objective function call. This makes the training more realistic in the sense that on a digital quantum computer, we will always have a finite sampling effect. All classical simulations are performed using the state vector simulation in Qiskit~\cite{javadiabhari2024quantumcomputingqiskit, Qiskit}, with no error model for noise other than the finite sampling effect. The readout bit measurements are mapped to spins as such: $0 \mapsto 1$ (up), $1 \mapsto -1$ (down). 

We use the blackbox parameter optimization algorithm Simultaneous Perturbation Stochastic Approximation (SPSA)~\cite{spall1998overview, 945806, 880982, 1220767} to train the VQE ansatz. This blackbox optimizer is a good general purpose technique to obtain close-to-optimal solutions of multi-variable objective function. This optimizer has specifically been used for training VQAs because of its robustness to finite sampling effects (shot noise), and generally performs better than other optimization algorithms~\cite{PhysRevA.108.032409, PhysRevA.107.032407, Kungurtsev2024iterationcomplexity}. To avoid prohibitively high costs associated with a large scale feedback loop between a classical optimizer and measuring the energy expectation value on a quantum computer, we optimize the energy in Eq.~\eqref{eq:costfuntion} on a classical computer with fixed number of shots (also known as pre-training).  For all VQE training iterations that we use in this study, the continuous VQE parameters are uniformly randomly initialized from $[0, 2\pi]$. For SPSA training, $40$ different random parameter initializations are executed with a maximum of $2000$ iterations per run.  Once a good estimate of global minimum Hamiltonian expectation value (energy) is found by VQE, we proceed to evaluate other necessary observables. The number of objective function calls is not necessarily equal to the number of optimizer iterations - optimizer iterations are defined specifically for each algorithm, and can require many objective function evaluations for a single optimizer iteration. Appendix~\ref{section:appendix_VQE_learning} details an example VQE learning iteration. The key observation from these VQE learning iterations is that the optimizer can become stuck in local minima, thus necessitating many random parameter initializations. 

The key to evaluating the accuracy of the VQE simulations is to compare it with exact results of the Hamiltonian. This is accomplished by using exact diagonalization of the Hamiltonian. However, this method is challenging to scale and requires High Performance Computing (HPC). Exact diagonalization of all $16$-site (qubits) Hamiltonians are computed through LAPACK based routines in the Python programming language~\cite{lapackcite}.

\textbf{Observables.} The first derivative of energy in Eq.~\eqref{eq:Hamiltonian}, although contains the information for the possible quantum phase transitions and their types (continuous or discontinuous), is generally not sufficient to probe the physical characteristics of individual phases. Two-site spin correlations contains necessary information about the particular phases. Moreover, one can use the spin-structure factor, which is the Fourier transformation of spin-spin correlations to identify the possible phases in the system. The correlation function and structure factor in our case is given as
\begin{align}
\label{eq:spin-structure}
\mathcal{C}_{i,j}=\langle \mathbf{S}_{i}\cdot\mathbf{S}_{j}\rangle\;,\quad\chi(\mathbf{q})=\frac{1}{n_x n_y}\sum_{j}e^{-i\mathbf{q}.\mathbf{r}_{j}}\mathcal{C}_{0,j} \;,
\end{align}

$\mathcal{C}_{i,j}$ is a two-spin correlation expectation value between spin at site $i$ ($\mathbf{S}_i=(X_i,Z_i)$) and spin at site $j$ with the indices taking values $(0,1,...,N-1)$. $\chi(\mathbf{q})$ is the spin-spin structure factor which is the 2D Fourier transformation of the sum of the spin-spin correlation functions between site 0 and sites $(0,1,...,N-1)$. $\mathbf{r}_{j}=x_j\hat{x}+y_j\hat{y}$ is the coordinate of the $j$-th site on the lattice. $\mathbf{q}=(q_x,q_y)$ is the wave vector where $q_x$ and $q_y$ take discrete values between $0,1,...,3$ in units of $2\pi/4$ for the $4\times 4$ square lattice. We use Eq.~\eqref{eq:spin-structure} to quantify the ability of the trapped-ion quantum computer to identify phases in our model as $J_2/J_1$ is changed. For each $(J_2,B_x)$ value, the observables like energy and two-spin correlators from the trained VQE circuits are evaluated using $10^6$ shots for each Pauli basis. This prevents additional finite sampling effect noise. Agreement with the ground-state energy is the key quantity that is measured for quantifying VQE accuracy. We also quantify the two correlators that make up Eq.~\eqref{eq:Hamiltonian} - the sum of spin-spin products for nearest-neighbor interactions and the sum for next-nearest-neighbor interactions which is also the first derivative of energy. In addition to Eq.~\eqref{eq:spin-structure}, observables like net transverse field $M_x=\sum_iX_i$ are also useful.

\begin{figure*}[th!]
\centering
\includegraphics[width=1.0\textwidth]{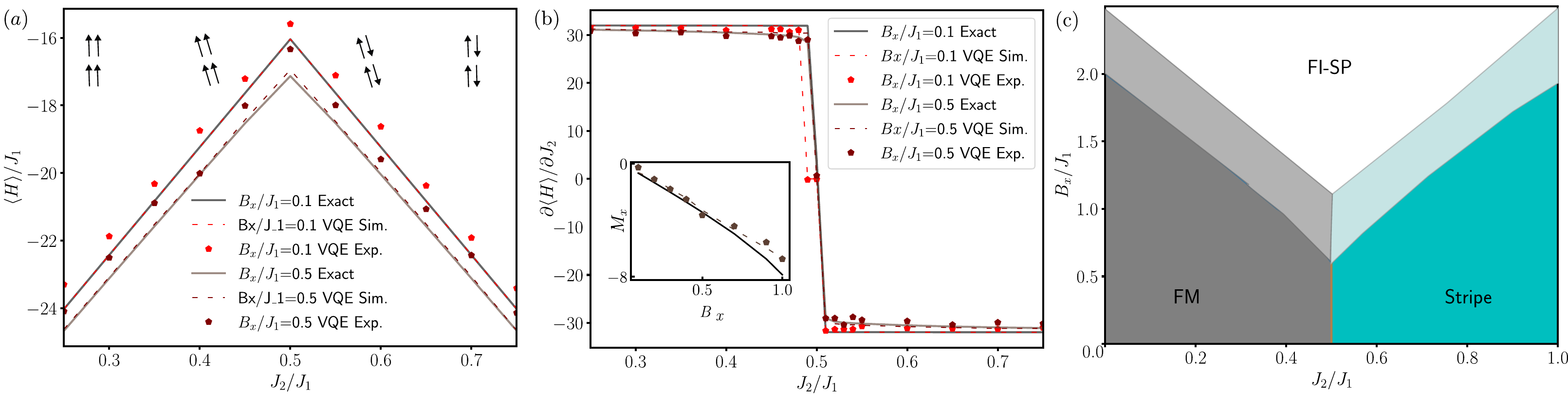}
\caption{(a) The ground-state energy profile of Eq.~\eqref{eq:Hamiltonian} for $N = 16$ under periodic boundary conditions obtained via VQE classical simulation, VQE H1-1 experiments (500 shots), and exact diagonalization. The exact-diagonalization solution is given as solid grey (brown) lines for $B_x/J_1=0.1$ ($B_x/J_1=0.5$). Classically computed VQE expectation values are given by red (brown) dashed lines. H1-1 computed expectation values are given as red (brown) pentagons for $B_x/J_1=0.1$ ($B_x/J_1=0.5$). The sketch at the top of (a) shows two distinct magnetic phases: ferromagnetic and stripe. (b) The derivative of the energy obtained through exact diagonalizationon, classical VQE simulator results, and H1-1 quantum processor using 100 shots showing the first-order phase transition. Inset gives the expectation value of transverse field magnetization $\sum_i X_i$ with respect to $B_x$ for $J_2/J_1=0.4$. The solid line, dashed line, and pentagons in the inset give exact, VQE simulation and experimental results for $M_x$ respectively. (c) Exact-diagonalization phase diagram showing FM, Stripe, and FI-SP phases. Lightly shaded areas represent co-existence regions where the spin structure factor shows peaks in both x- and z- directions.}
\label{fig:observables}
\end{figure*}

\begin{figure}[th!]
\centering
\includegraphics[width=0.49\textwidth]{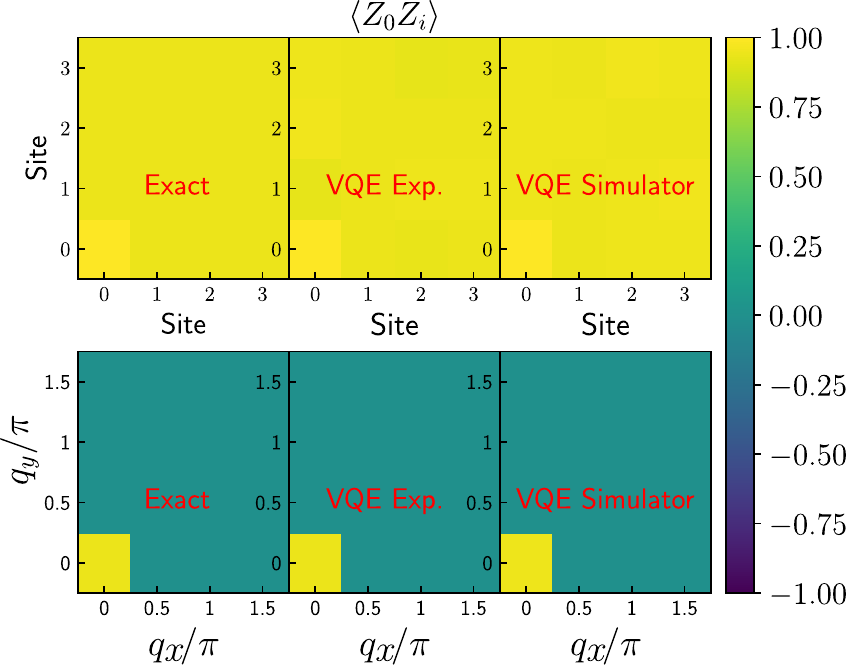}
\caption{ The comparison of exact, VQE experiment results on Quantinuum H1-1 trapped-ion processor (with 500 shots), and VQE classical simulator for $B_x/J_1 = 0.5$ and $J_2/J_1=0.45$. The top row gives the comparison of spin-spin correlation function $\langle Z_0 Z_i\rangle$. The bottom row gives the spin structure factor obtained from the correlation function showing the presence of ferromagnetic instability.  }
\label{fig:FM_insability}
\end{figure}

\begin{figure}[th!]
\centering
\includegraphics[width=0.49\textwidth]{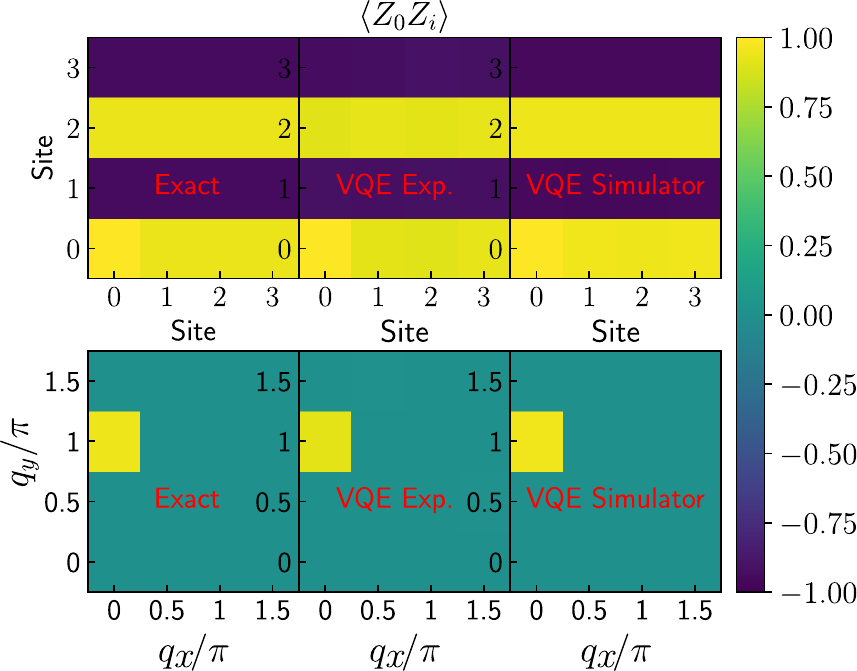}
\caption{ The comparison of exact, VQE classical simulator and VQE experiment results on Quantionuum H1-1 trapped-ion quantum processor (with 500 shots) for $B_x/J_1 = 0.5$ and $J_2/J_1=0.55$. The spin-spin correlation (top row) and the spin structure factor (bottom row) shows the presence of $(0,\pi)$ (stripe-phase) instability. Exact-diagonalization calculations are performed with small a staggered field (field's direction alternates with each row) to remove the possible superpositions of different degenerate stripe patterns. }
\label{fig:SP_insability}
\end{figure}

%%%%%%%%%%%%%%%%%%%%%%%%%%%%%%%%%%%%%%%%%%%%%%%%%%%%%%%%%%%%%%%%%%%%%%%%%%%%%%%%%%%%%%%%
\section{Results and Discussions}
\label{section:results}
%%%%%%%%%%%%%%%%%%%%%%%%%%%%%%%%%%%%%%%%%%%%%%%%%%%%%%%%%%%%%%%%%%%%%%%%%%%%%%%%%%%%%%%%
Using the best found classically trained VQE parameters, outlined in Sec.~\ref{section:methods}, the VQE circuits are executed on the Quantinuum H1-1 processor. We performed quantum simulation of the model with transverse field strengths between $0.1-1$ in units of $J_1$. The usual values of $J_1$ is in the order of 10meV which is equivalent of a magnetic field of about 100 Tesla, still higher than experimentally accessible fields. Fig.~\ref{fig:observables}(a) shows the ground-state energy profile of our model for various values of frustration parameter $J_2/J_1$ for $B_{x}/J_1=0.1,0.5$; there is nearly a perfect agreement between exact and VQE classical simulator results. The VQE experiment results in Fig.~\ref{fig:observables}(a) reproduce the ground-state energy profiles for both of the transverse-field runs without using any error-mitigation methods on the H1-1 quantum computer (500 shots). Earlier work~\cite{prrbenedikt} failed to reproduce the correct qualitative energy profile for a similar Ising model in 1D when evaluated from quantum experiments even after the application of error mitigation methods. Fig.~\ref{fig:observables}(b) gives the first derivative of energy but the experimental measurements are based on 100 quantum circuit executions (shots). The experimental results show a step transition between different phases as the frustration parameter increases, agreeing with the exact results. Fig.~\ref{fig:observables}(c) shows the expected phase diagram obtained by exact diagonalization.

\begin{figure}[th]
\centering
\includegraphics[width=0.45\textwidth]{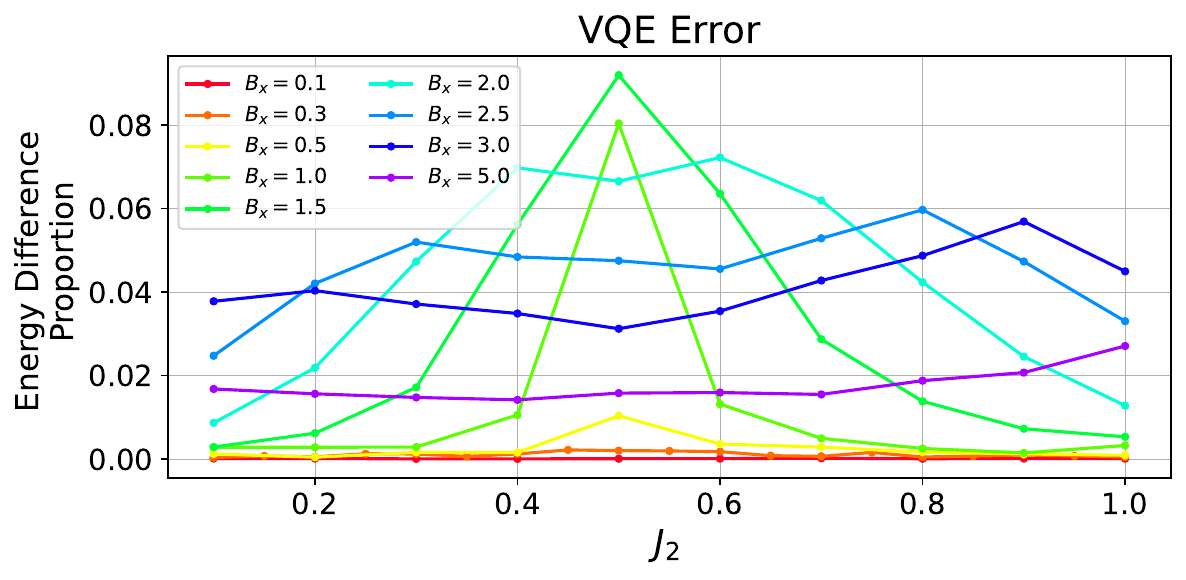}
\caption{SPSA trained VQE energy compared to the true ground state energy, computed using exact diagonalization. The quantity being plotted on the y-axis is error between the exact energy and the (best found) VQE energy, as a proportion (minimum error is $0$, maximum is $1$). The x-axis is the Hamiltonian frustration parameter $J_2$. The VQE gives very low error rates at small $B_x$ values, and begins to achieve low error rates at very high $B_x$ well into the paramagnetic region ($B_x=5$). However, close to $J_2=0.5$ and more intermediate $B_x$ values the VQE error rate becomes higher. VQE expectation values evaluated with $10^6$ samples per Pauli basis. }
\label{fig:VQE_exact_diagonalization_error_SPSA}
\end{figure}

Other quantities that encode information about the ground-state phases are the spin-spin correlation function and the spin structure factor given in Eq.~\eqref{eq:spin-structure}. Fig.~\ref{fig:FM_insability} shows the correlation function $\langle Z_0 Z_i \rangle$ and the structure factor for $B_x/J_1=0.5$ and $J_2/J_1=0.45$ for ferromagnetic instability. We characterize ferromagnetic instability when a spin-spin correlator between site 0 and other sites gives $\langle Z_0Z_i\rangle\to 1$. Moreover, the structure factor which is the 2D Fourier transform of the correlation function will show a strong peak at $\mathbf{q}=0$ for the ferromagnetic phase. The correlation function and the structure factor obtained from exact results, VQE classical simulator results, and the VQE H1-1 experimental data (with 500 shots) show perfect agreement. Similarly, a stripe-phase instability, or $(0,\pi)$ order, is clearly identified in Fig.~\ref{fig:SP_insability} for $B_x/J_1=0.5$ and setting $J_2/J1=0.55$. In this phase, spins in the adjacent rows (columns) anti-align giving us stripe pattern in the spin-spin correlation function. The structure factor will then show peak(s) at $\mathbf{q}=(0,\pi)$ or/and $\mathbf{q}=(\pi,0)$. For this phase, the H1-1 signal with 500 shots also agrees with VQE and exact results shown in Fig.~\ref{fig:SP_insability}. We have also simulated the spin-spin correlator for the $x$-basis component (not shown) of the spin and find a consistent conclusion. The values of $J_2/J_1$ in Figs.~\ref{fig:FM_insability} and~\ref{fig:SP_insability} are chosen near the frustration point.

The phase diagram of the TFIM with ferromagnetic-nearest-neighbor and antiferromagnetic-next-nearest-neighbor interactions has been presented in Ref.~\onlinecite{Oitmaa_2020_TFANNNI}. By using series-expansion methods, the phase boundaries were located to high precision. The work showed that for $J_2/J_1<0.5$ and values up to $B_x/J_1\leq2$, the system exhibits ferromagnetic instability. While for $J_2/J_1>0.5$, the system is in a stripe ($(0,\pi)$ or $(\pi,0)$) ordered phase for $B_x/J_1\leq2$. For the larger values of $B_x$ the system is in a field-induced spin-polarized (FI-SP) state sometimes called $\Gamma$- or paramagnetic phase. The work also predicts that there is a 1st-order phase transition between the ferromagnetic and stripe phase at the frustration point $J_2/J_1=0.5$ for values of $B_x$ up to $2J_1$. We found that the 1st-order phase transition between ferromagnetic and stripe phase is only present upto to $B_x/J_1=0.8$ in our exact-diagonalization calculations. This can be due to finite size effects and coexistence of different phases as $B_x$ is increased from $0$. In Fig.~\ref{fig:observables}(c), we present our phase diagram through computing the exact ground state. We identify the phases as FM phase if the structure factor obtained from spin-spin correlation function in Z-basis peaks at $\mathbf{q}=0$ and the structure factor for the correlations in the x- direction does not show any strong peak. Similarly, the stripe-phase location is obtained through structure factor of spin-spin correlator along the z-axis showing peak(s) at $\mathbf{q}=(0,\pi)$ or $\mathbf{q}=(\pi,0)$ or the superposition of both. When the structure factor of the spin-spin correlator along the x-direction shows a strong peak at $\mathbf{q}=0$, we label the state as F1-SP state, since in this state the spins are aligning along the x-axis due to the high transverse field. The lightly shaded areas in Fig.~\ref{fig:observables}(c) are obtained when the structure factor shows peaks in both z- and x-directions. This indicates coexistence regions between different phases and should vanish in the thermodynamical limit.

Fig.~\ref{fig:VQE_exact_diagonalization_error_SPSA} quantifies the error between the lowest-energy state found from extensive SPSA classical training of the VQE ansatz and the exact ground state energy, across a wide range of $B_x$ and $J_2$ values. This VQE error is estimated, post-training, using $10^6$ shots per Hamiltonian parameter with the goal of minimizing the effect of shot noise on this quantity. This quantity is error proportion defined as $\frac{s_e - E_{GS}}{|E_{GS}|}$, where $E_{GS}$ is the ground-state energy and $s_e$ is the lowest found energy from VQE. Therefore in Fig.~\ref{fig:VQE_exact_diagonalization_error_SPSA}, when this quantity is $0$ that means VQE perfectly found the true ground-state energy, and the maximum possible error quantity is $1$. Fig.~\ref{fig:VQE_exact_diagonalization_error_SPSA} shows that there is a maximum VQE error of $\approx 0.08$ that occurs near $J_2=0.5$ for $B_x=1.0-1.5$. In the regime where the VQE error is highest, reaching up to $\approx 8\%$, the exact mechanism for this highest error of the VQE simulation is not known. This region of the phase diagram does contain a significant amount of  frustration and quantum fluctuations. Therefore, it is plausible that VQE has increased difficulty learning the ground-state in this region of the phase diagram. In particular, we believe that the most likely explanation is that the classical learning optimizer becomes stuck in sub-optimal local minima. This is evidenced by the fact that SPSA optimizer becomes stuck in local minima in several instances, where we observe other random VQE parameter initializations performing better (see Appendix~\ref{section:appendix_VQE_learning}). Therefore, for the higher-error data points in Fig.~\ref{fig:VQE_exact_diagonalization_error_SPSA} all SPSA learning iterations we performed simply could not finding the global minimum.

%%%%%%%%%%%%%%%%%%%%%%%%%%%%%%%%%%%%%%%%%%%%%%%%%%%%%%%%%%%%%%%%%%%%%%%%%%%%%%%%%%%%%%%%
\section{Conclusion}
\label{section:conclusion}
%%%%%%%%%%%%%%%%%%%%%%%%%%%%%%%%%%%%%%%%%%%%%%%%%%%%%%%%%%%%%%%%%%%%%%%%%%%%%%%%%%%%%%%%

We used variational and exact treatments to obtain the ground-state profile of 2D TFIM with frustration to benchmark the experimental signal from a trapped-ion Quantinuum computer. For our Variational Quantum Algorithm, we proposed a hardware efficient VQE ansatz where two-qubit entangling gates mirror nearest-neighbor and next-nearest-neighbor Ising-exchange interactions. The VQE was then trained classically in the presence of shot noise to obtain an upper bound on the ground-state energy for varying transverse field strengths and frustration parameters. These trained VQE circuits were then run on the Quantinuum H1-1 trapped-ion device. The quantum computer experimental signal agrees with exact diagonalization for various quantities including energy, the first derivative of energy, two-spin correlation functions and the spin structure factor. Hence, we were able to map out the ground-state phases of the system even in the low shot (100-500) limit. In stark contrast to other superconducting-device experiments, we did not use quantum error-mitigation methods to obtain the ground-state phases. Such high-quality results were difficult to obtain on superconducting-qubit quantum devices even for 1D systems and using high shot counts~\cite{prrbenedikt}. Thus, current trapped-ion devices appear to be the most promising quantum platform to study properties of strongly correlated systems. As it turns out, the all-to-all connectivity of trapped-ion devices is a key feature that enables the study of 2D models with competing interactions. 

%%%%%%%%%%%%%%%%%%%%%%%%%%%%%%%%%%%%%%%%%%%%%%%%%%%%%%%%%%%%%%%%%%%%%%%%%%%%%%%%%%%%%%%%
\section*{Acknowledgments}
\label{sec:acknowledgments}
%%%%%%%%%%%%%%%%%%%%%%%%%%%%%%%%%%%%%%%%%%%%%%%%%%%%%%%%%%%%%%%%%%%%%%%%%%%%%%%%%%%%%%%%

E.P. thanks Victor Drouin-Touchette for helpful discussions. A.K. and J.-X.Z. acknowledge helpful discussion with Benedikt Fauseweh. This work was carried out under the auspices of the U.S. Department of Energy (DOE) National Nuclear Security Administration (NNSA) under Contract No. 89233218CNA000001. It was supported by LANL LDRD Program under project number 20230049DR  (A.K.), the NNSA Advanced Simulation and Computing Beyond Moore's Law Program (E.P., A.B. \& S.E.) and Physics and Engineering Models (J.-X.Z.), Quantum Science Center, a U.S. DOE Office of Science Quantum Information Science Research Center (A.K.), and in part by Center for Integrated nanotechnologies, a DOE BES user facility, in partnership with the LANL Institutional Computing Program for computational resources. The classical simulations part of the results used the Darwin testbed at Los Alamos National Laboratory, which is funded by the Computational Systems and Software Environments subprogram of LANL Advanced Simulation and Computing program (NNSA/DOE). We used quantum resources of the Oak Ridge Leadership Computing Facility, which is a DOE Office of Science User Facility supported under Contract DE-AC05-00OR22725.
\appendix

%%%%%%%%%%%%%%%%%%%%%%%%%%%%%%%%%%%%%%%%%%%%%%%%%%%%%%%%%%%%%%%%%%%%%%%%%%%%%%%%%%%%%%%%
\section{Measured Probability Based VQE Phase Diagram}
\label{section:appendix_VQE_phase_diagram}
%%%%%%%%%%%%%%%%%%%%%%%%%%%%%%%%%%%%%%%%%%%%%%%%%%%%%%%%%%%%%%%%%%%%%%%%%%%%%%%%%%%%%%%%
\begin{figure}[th!]
\centering
\includegraphics[width=0.45\textwidth]{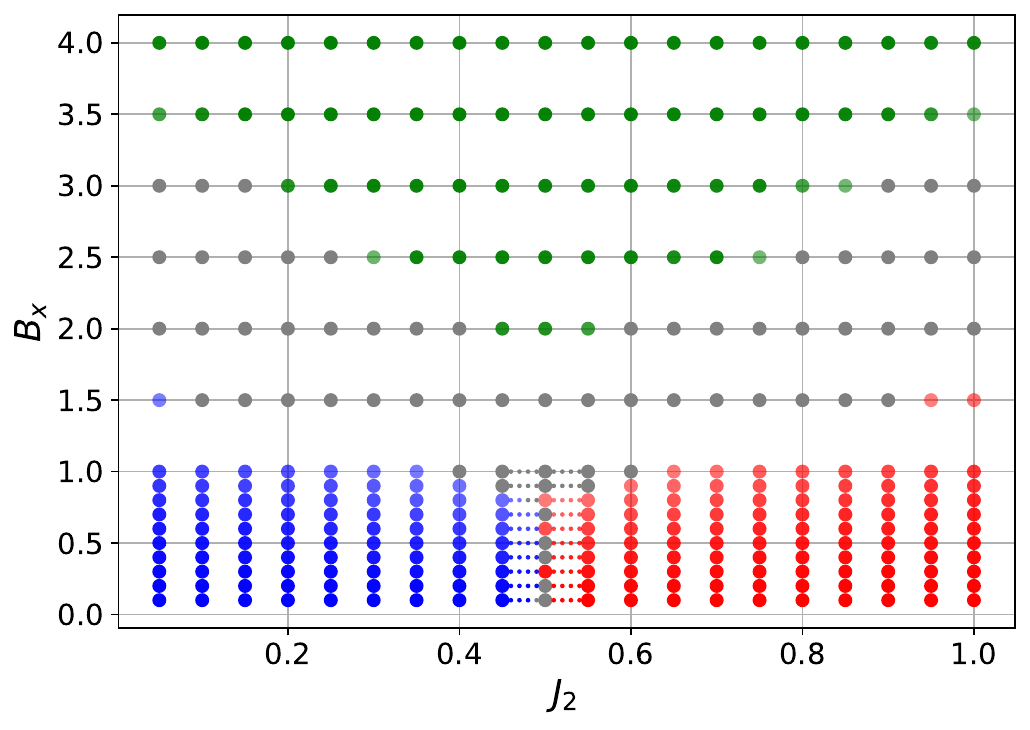}
\caption{VQE-simulator computed phase diagram on a grid of Hamiltonian parameters $(J_2,B_x)$ with finite sampling error. Blue denotes that the total measured probabilities of the two possible ferromagnetic ground states (all spin up and all spin down) in the $Z$-basis is greater than $0.5$. Red denotes that the total measured probability of the four degenerate stripe states in the $Z$-basis is greater than $0.5$. Green denotes that the total measured probability of the $\ket{+}^n$ (FI-SP) state from the $X$-basis measurements is greater than $0.5$. All states which do not fall into either of these three categories are colored as grey, denoting regions where there is not a dominant magnetic phase as found by the VQE simulation. Transparency of the coloring of the points equals the total measured probabilities for the paramagnetic, ferromagnetic, and antiferromagnetic measured probabilities respectively. Each point in this phase diagram (for each Pauli basis measurement) were evaluated using a total of $1\mathrm{e}{6}$ shots. Higher resolution of $(J_2,B_x)$ near $J_2=0.5$ and below $B_x=1.0$ verifies VQE performance in this frustrated region (the points are smaller in size to accommodate the rest of the points on the diagram).}
\label{fig:VQE_phase_diagram}
\end{figure}
Fig.~\ref{fig:VQE_phase_diagram} shows phase diagram obtained by classifying the magnetic phases from classically trained VQE circuits. Phases are identified by measuring the probabilities for the different magnetic orders in both the Pauli X and Z basis and selecting the highest probability magnetic order. This is a point-wise phase diagram, with no extrapolation used to demark the different magnetic phases. At $J_2=0.5$, VQE at different values of $B_x$ can give stripe or ferromagnetic phase or no dominant phase (denoted as grey points). It is harder for the VQE to approximate the ground-state pattern, since there is large degeneracy at the frustration point. The grey region seen in Fig.~\ref{fig:VQE_phase_diagram} matches the coexistence regions found in Fig.~\ref{fig:observables}-(c) by exact diagonalization.

%%%%%%%%%%%%%%%%%%%%%%%%%%%%%%%%%%%%%%%%%%%%%%%%%%%%%%%%%%%%%%%%%%%%%%%%%%%%%%%%%%%%%%%%
\section{Complete 2D TFIM Observables From VQE Simulations}
\label{section:appendix_more_VQE_observables}
%%%%%%%%%%%%%%%%%%%%%%%%%%%%%%%%%%%%%%%%%%%%%%%%%%%%%%%%%%%%%%%%%%%%%%%%%%%%%%%%%%%%%%%%
\begin{figure}[th!]
\centering
\includegraphics[width=0.49\textwidth]{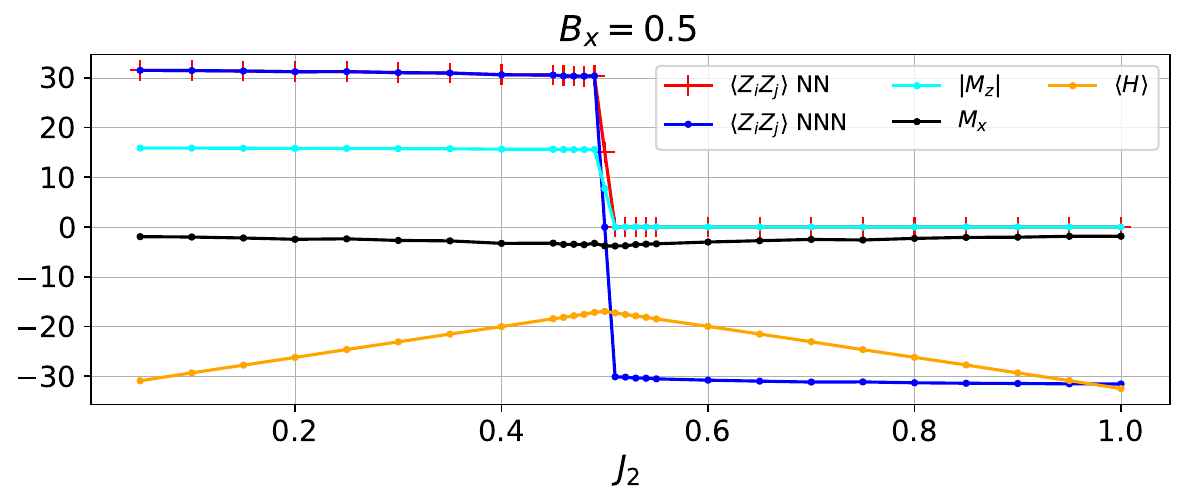}
\includegraphics[width=0.49\textwidth]{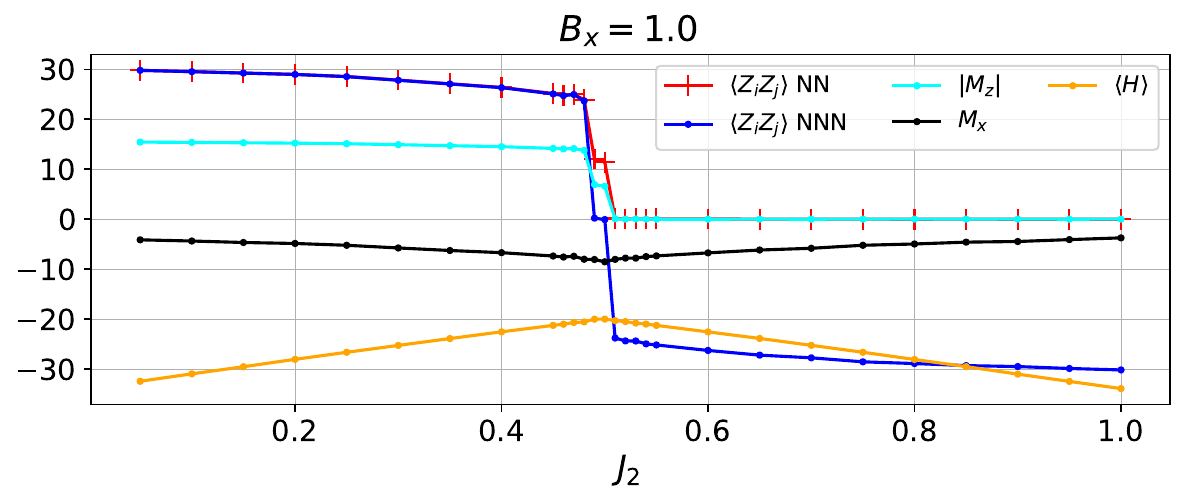}
\includegraphics[width=0.49\textwidth]{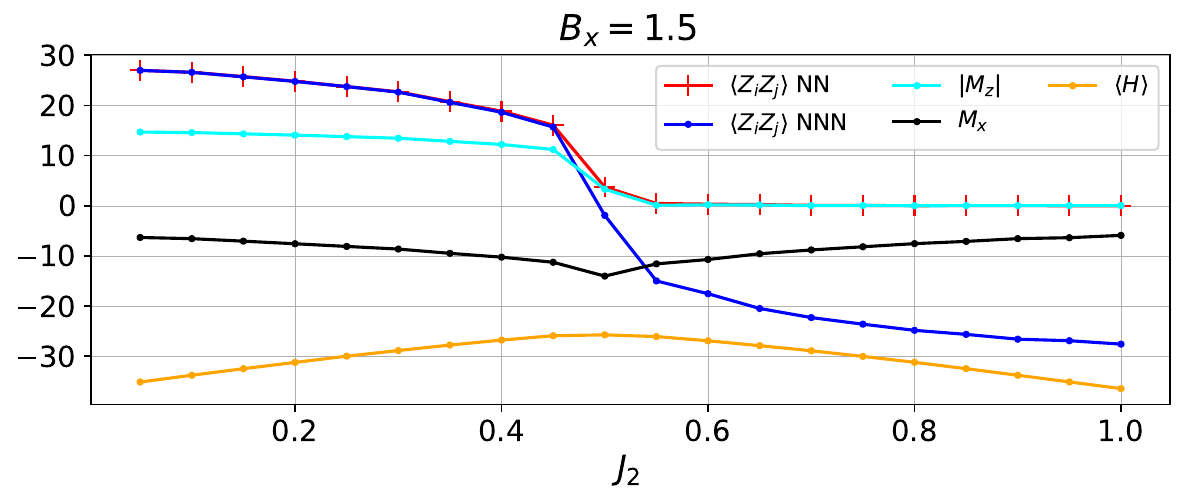}
\includegraphics[width=0.49\textwidth]{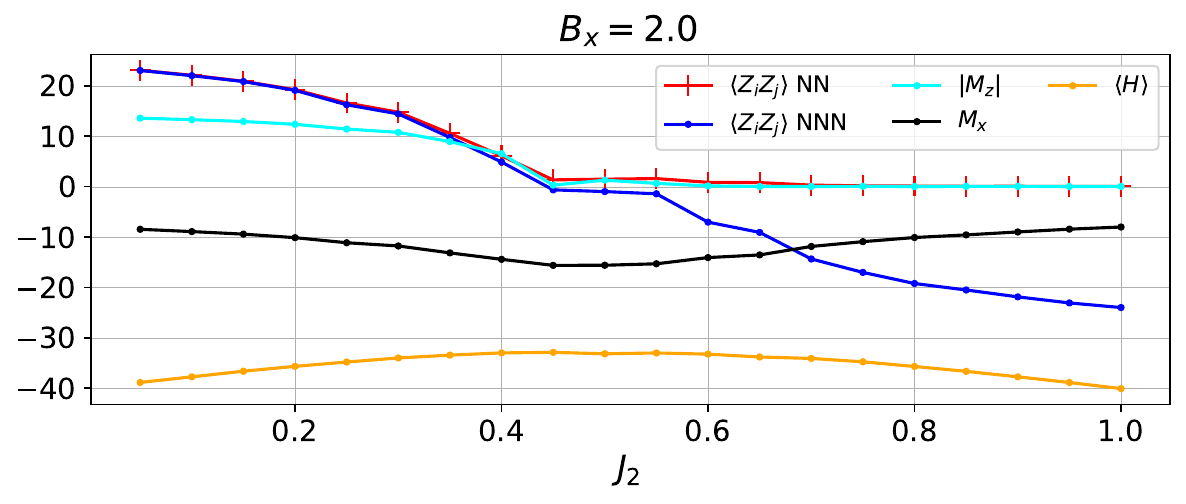}
\caption{Observables as a function of $J_2$ from trained VQE simulations for different $B_x$. Each sub-plot shows the computed observables given by the parameters corresponding to the lowest Hamiltonian expectation value in the classical optimization. These observables are the expectation values (computed with $10^6$ shots per parameter per Pauli basis) of net magnetization in the Z-basis measurements (plotted as the absolute value of $M_z$), net magnetization in the X-basis $M_x$, nearest neighbor spin-spin correlations ($\langle Z_i Z_j \rangle$ NN), next-nearest-neighbor spin-spin correlations ($\langle Z_i Z_j \rangle$ NNN), and energy ($\langle H \rangle$). The shared y-axis simultaneously denotes the expectation values of each of these observables. }
\label{fig:additional_VQE_observables}
\end{figure}
Fig.~\ref{fig:additional_VQE_observables} plots the quantities of several different observables as computed by VQE, as a function of $J_2$, for several different $B_x$ values upto $B_x=2$. All of the VQE simulation data shown in this figure comes from exact classical simulations of the VQE circuits with no noise model. We used the optimized VQE parameters found for each one of these Hamiltonian parameters. These plots show the expection values of different terms within the Hamiltonian that is being simulated by VQE. At low $B_x$, the observables, in particular the spin-spin correlations, change steeply at $J_2=0.5$ whereas for higher $B_X$ the observables change more gradually and get more flat as a function of $J_2$. This shows that classical VQE simulator results can differentiate between step and continuous phase transitions.

%%%%%%%%%%%%%%%%%%%%%%%%%%%%%%%%%%%%%%%%%%%%%%%%%%%%%%%%%%%%%%%%%%%%%%%%%%%%%%%%%%%%%%%%
\section{1D TFIM with competing nearest neighbour and next-nearest neighbour interaction VQE Simulations on the Quantinuum H1-1 processor}
\label{section:appendix_1D_Quantinuum_emulator}
%%%%%%%%%%%%%%%%%%%%%%%%%%%%%%%%%%%%%%%%%%%%%%%%%%%%%%%%%%%%%%%%%%%%%%%%%%%%%%%%%%%%%%%%
\begin{figure}[th!]
\centering
\includegraphics[width=0.49\textwidth]{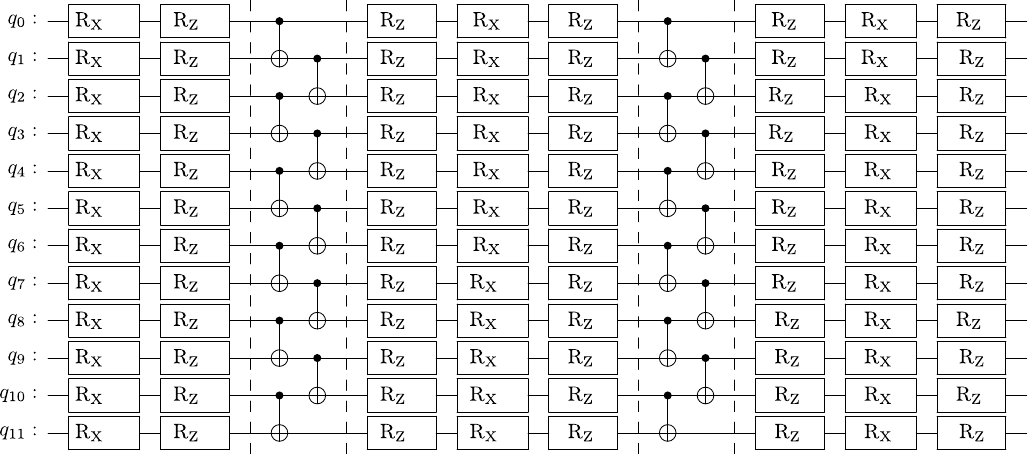}
\caption{Linear-nearest-neighbors connectivity hardware-efficient VQE ansatz for 1D TFIM simulation with $J_1-J_2$ frustrating interactions. Each single qubit gate is parameterized by a different continuous parameter in the interval from $[0, 2\pi]$. This ansatz is used in the simulations shown in Fig.~\ref{fig:PRR1d}(taken from~\cite{prrbenedikt}). }
\label{fig:1dcircuit}
\end{figure}

\begin{figure}[th!]
\centering
\includegraphics[width=0.45\textwidth]{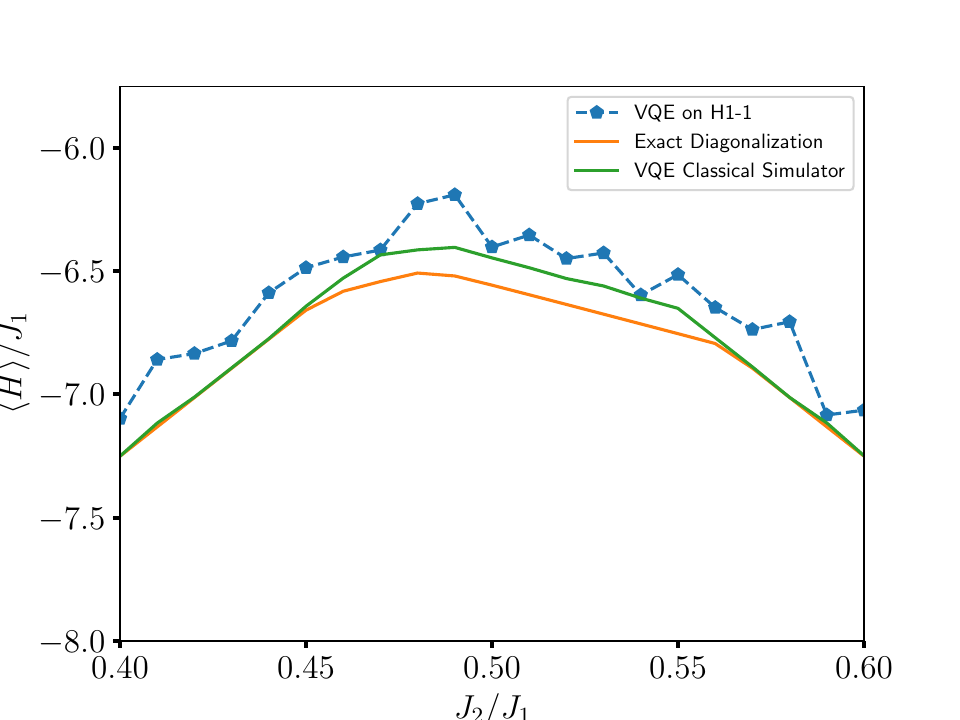}
\caption{Ground state energy as a function of $J_2$ for the 1D $12$-qubit TFIM. H1-1 results are obtained with 100 shots. ``VQE Classical Simulator'' denotes the expectation value given by classical simulation with finite shot noise, and no error model. ``Exact Diagonalization" denotes the expectation value given by exact diagonalization. These simulations used the ansatz given in Fig.~\ref{fig:1dcircuit}. This plot follows the methods and VQE ansatz given in ref. \cite{prrbenedikt}, but is now being simulated on the H1-1 noisy emulator. }
\label{fig:PRR1d}
\end{figure}

To briefly compare the VQE simulations of 1D $J_1-J_2$ frunstrated TFI model on H1-1 processor with the 1D results of superconducting quantum processors given in ~\cite{prrbenedikt}, we repeat the calculations of~\cite{prrbenedikt} for a periodic 1D lattice of size 12. To this end, here we follow the VQE training methodology described in~\cite{prrbenedikt}, which is different to what was done for the 2D system VQE simulations. We first optimize (maximize) the state overlap between a parameterized statevector object and the exact-wavefunction, in Qiskit~\cite{Qiskit, javadiabhari2024quantumcomputingqiskit} using the dual-annealing optimizer for few thousand iteration. The output of this optimization is then used as an initial guess for VQE classical simulations using the SPSA optimizer for 100,000 iterations in the presence of finite shot noise giving us classically-precomputed VQE parameters. The VQE ansatz that is used is given in Fig.~\ref{fig:1dcircuit}. Once good agreement with the exact ground state energy is reached using the classical optimizer, these precomputed VQE parameters are then run on the Quantinuum H1-1 trapped-ion device. The comparison between exact diagonalization, classically computed VQE, and H1-1 results are shown in Fig.~\ref{fig:PRR1d}. Fig.~\ref{fig:PRR1d} shows the energy-profile of the 1D frustrated TFI model, using no error mitigation on the Quantinuum H1-1 trapped-ion device. The energy profile agrees with the exact solution and the result is qualitatively good even with 100 shots.

%%%%%%%%%%%%%%%%%%%%%%%%%%%%%%%%%%%%%%%%%%%%%%%%%%%%%%%%%%%%%%%%%%%%%%%%%%%%%%%%%%%%%%%%
\section{VQE Learning Curve from SPSA}
\label{section:appendix_VQE_learning}
%%%%%%%%%%%%%%%%%%%%%%%%%%%%%%%%%%%%%%%%%%%%%%%%%%%%%%%%%%%%%%%%%%%%%%%%%%%%%%%%%%%%%%%%
\begin{figure}[th!]
\centering
\includegraphics[width=0.45\textwidth]{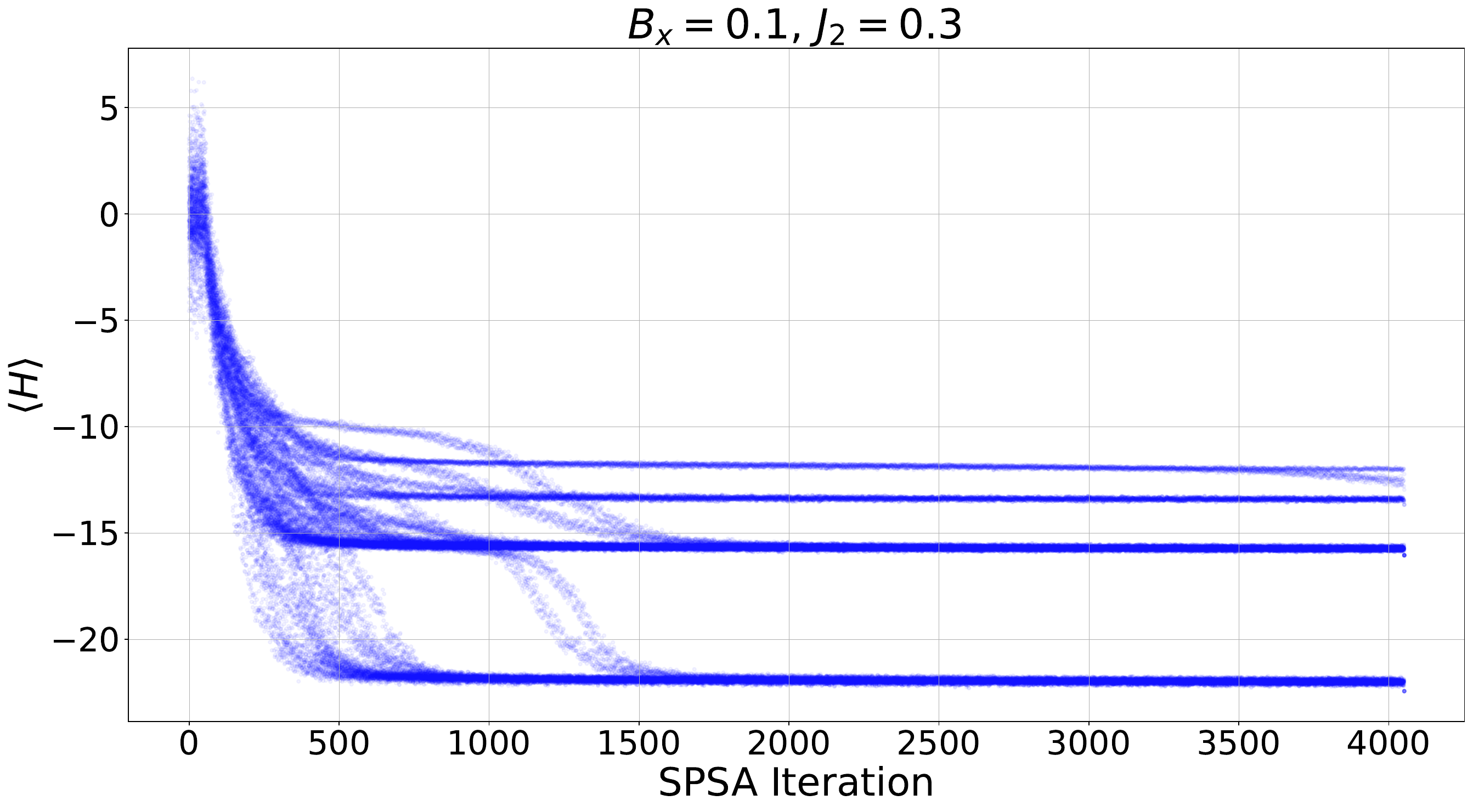}
\caption{Example VQE learning convergence using the SPSA optimizer. This plot contains the energy learning curves of $40$ different learning initializations, where for each run all VQE single qubit tunable parameters are uniformly randomly initialized from the interval $[0, 2\pi]$. The y-axis is the cost function value, e.g., the estimated expectation value of the Hamiltonian. Hamiltonian parameters of $B_x=0.1$ and $J_2 = 0.3$. The total number of SPSA iterations is always set to $2000$, but the actual number of cost function evaluations is on the order of 2 times the number of iterations (x-axis). Here each independent learning curve is represented by plotting the evaluated energy (cost function value) as a function of the learning iteration, where each evaluated expectation value is shown as a high transparency point with no interpolation between points. This learning curve shows that the effect of of the initial VQE parameter choice significantly impacts the performance of the optimizer and thereby the quality of the VQE simulation, and in particular SPSA can get stuck in local minima depending on this initial parameter choice. Namely, despite all of these learning runs having (qualitatively) converged by approximately $1500$ iterations, independently initialized runs did not converge to the same expectation value.  }
\label{fig:VQE_learning_SPSA}
\end{figure}

Fig.~\ref{fig:VQE_learning_SPSA} shows an example set of VQE learning iterations using the SPSA optimizer. This plot shows that it is necessary to use many random VQE parameter initializations in order for the VQE learning to perform well and not get stuck in local minima as frequently -- in this case, and for all Hamiltonian parameters used in this study, $40$ random initializations are used. This plot serves as a representative example of the VQE parameter tuning performed with SPSA, and it suggests that one of the primary limitations of the VQE for this model is the capability of the black-box optimizer.

\section{Periodic Boundary Condition on a Torus}
\label{section:appendix_periodic_boundary_condition}
%% TIKZ
\begin{figure}[ht]
    \centering
    %\hfill%
        \begin{tikzpicture}[scale=1.5]
        \tikzmath{
            \n = 4;
            \m = 4;
        }
        %% iterate over tuples (x,y)
        %% calculate    idx = x*n + y
        \foreach \x in {0,...,\inteval{\n-1}}
            \foreach \y in {0,...,\inteval{\m-1}}
            {
                \tikzmath{
                    int \idx, \diagshade, \andgshade;
                    \idx = {\y*\n+\x};
                    \diagshade = {50*(1+Mod(\x-\y,gcd(\n,\m))/(gcd(\n,\m)-1))};
                    \andgshade = {50*(1+Mod(\x+\y,gcd(\n,\m))/(gcd(\n,\m)-1))};
                }
                %% node
                \node at (\x,\y) {\scriptsize $\idx$};
                %% edges
                \draw[lightgray,very thin,bend left] (\x,\y) edge ({Mod(\x+1,\n)},\y);                          % horizontal:   (x, y) -> (x+1 mod n, y)
                \draw[gray,very thin,bend left] (\x,\y) edge (\x,{Mod(\y+1,\m)});                               % vertical:     (x, y) -> (x, y+1 mod m)
                \draw[red!\diagshade!black,very thin,bend left] (\x,\y) edge ({Mod(\x+1,\n)},{Mod(\y+1,\m)});   % diagonal:     (x, y) -> (x+1 mod n, y+1 mod m)
                \draw[blue!\andgshade!white,very thin,bend left] (\x,\y) edge ({Mod(\x+1,\n)},{Mod(\y-1,\m)});  % anti-diagonal:(x, y) -> (x+1 mod n, y-1 mod m)
            } 
    \end{tikzpicture}
    \caption{Node connectivity graphs with toroidal periodic boundary conditions on a square lattice $n=m=4$. Gray edges denote horizontal and vertical edges, anti-diagonal edges are blue, and diagonal edges are red. }
    \label{fig:Toroidal_boundary_condition_graph_structure}
\end{figure}
To go closer to the thermodynamic limit and to prevent unwanted boundary effects, it is preferable to use periodic boundary conditions of the 2D model. In particular, we use toroidal boundary conditions. Here we outline the convention we use to the create toroidal boundary conditions based on the original model's 2D lattice. For a grid of length $n$ and height $m$, and grid points $(x,y)$ with $0\leq x \leq n,\ 0\leq y\leq m$, we use the following indexing convention:
\[ (x,y) \longmapsto y\cdot n + x \]

\noindent
We then have \textbf{NN-connections}:
\begin{itemize}[noitemsep]
    \item horizontal edges: $(x, y)$ --- $(x+1 \bmod n, y)$\\
          These form $m$ many loops of length $n$.
    \item vertical edges:\ $(x, y)$ --- $(x, y+1 \bmod m)$\\
          These form $n$ many loops of length $m$.
\end{itemize}
and \textbf{NNN-connections}:
\begin{itemize}[noitemsep]
    \item diagonal edges: $(x, y)$ --- $(x+1 \bmod n, y+1 \bmod m)$\\
          These form $gcd(n,m)$ many loops of length $lcm(n,m)$. Here $gcd$ stands for greatest common divisor and $lcm$ stands for least common multiple. 
    \item anti-diagonal edges:\ $(x, y)$ --- $(x+1 \bmod n, y-1 \bmod m)$\\
          These form $gcd(n,m)$ many loops of length $lcm(n,m)$.
\end{itemize}

Fig.~\ref{fig:Toroidal_boundary_condition_graph_structure} renders the graph structure of these boundary conditions for a $4 \times 4$ system. Note that for $n=m$, the 4 types of each form $n$ many loops of length $n$. Note that this type of graph is also known as a kings graph (with periodic boundary conditions). These edges define the two qubit entangling gates used for the VQE ansatz.

\bibliography{references}

\end{document}